\documentclass[journal=jacsat,manuscript=article]{achemso}

\usepackage[T1]{fontenc}
\usepackage{chemformula}
\usepackage{hyperref}
\usepackage{graphicx}%
\usepackage{multirow}%
\usepackage{amsmath,amssymb,amsfonts}%
\usepackage{amsthm}%
\usepackage{mathrsfs}%
\usepackage[title]{appendix}%
\usepackage{xcolor}%
\usepackage{textcomp}%
\usepackage{manyfoot}%
\usepackage{booktabs}%
\usepackage{algorithm}%
\usepackage{algorithmicx}%
\usepackage{algpseudocode}%
\usepackage{listings}%

\usepackage[labelfont=bf]{caption}

\usepackage{graphicx}
\usepackage[leftcaption]{sidecap}

\usepackage[tight-spacing=true]{siunitx}
\usepackage[capitalise,nameinlink]{cleveref}

\author{Karol M. D\k{a}browski}
\affiliation[AGH University of Krakow]
{\footnotesize{Faculty of Drilling, Oil and Gas, AGH University of Krakow, al. Mickiewicza 30, 30-059, Krakow, Poland}}
\email{karol.dabrowski@agh.edu.pl}
\author{Mohammad Nooraiepour}
\affiliation[University of Oslo]{\footnotesize{Environmental Geosciences, Department of Geosciences, University of Oslo,1047 Blindern, Oslo, Norway}}
\author{Mohammad Masoudi}
\affiliation[SINTEF]{\footnotesize{SINTEF Industry, Applied Geoscience Department, 7465 Trondheim, Norway}}

\title[Salt Precipitation Dynamics in Geological CO$_2$ Storage]
  {\large{Quantifying Salt Precipitation During CO$_2$ Injection:\\ How Flow Rate, Temperature, and Phase State\\ Control Near-Wellbore Crystallization}}

\abbreviations{IR,NMR,UV}

\begin{document}
\begin{tocentry}
	\includegraphics[width=1\textwidth]{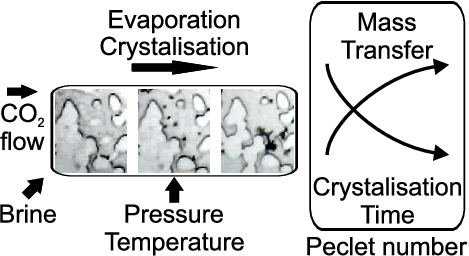}
\end{tocentry}

\begin{abstract}
\noindent Salt precipitation near injection wells can reduce permeability, induce excess pressure buildup, and reduce injectivity within days to weeks of CO$_2$ injection, yet the pore-scale mechanisms coupling multiphase flow, evaporation, and crystallization warrant further detailed quantification across variable phase states and flow regimes. We present high-resolution microfluidic experiments that systematically quantify the dynamics of halite crystallization during CO$_2$-driven brine evaporation across liquid, gaseous, and supercritical phases (50--80~bar, 20--60$^\circ$C, Pe = 50--1440). Crystallization kinetics are controlled by transport, with the Avrami rate constant ($K$) increasing by two orders of magnitude with the Péclet number and exhibiting the dependence of the temperature of Arrhenius ($E_a$ = 58.6~kJ ~ mol$^{-1}$). Supercritical CO$_2$ achieves superior displacement efficiency (residual saturation 0.22--0.36, fractal dimension $D$ = 1.79--1.82) and the fastest evaporation (Sherwood numbers 2--3$\times$ higher than the liquid phase), reducing the nucleation time from 57~min (20$^\circ$C, liquid) to $<$1~min (40--60 $^\circ$C, gas/supercritical). The final fractions of crystal increase 10-fold from liquid (0.008) to gas-phase conditions (0.08--0.12), confirming that convective transport and phase state dominate over diffusion-limited mechanisms. Despite probabilistic nucleation, final crystal distributions are spatially rather uniform with no systematic inlet-outlet bias. These quantitative relationships between dimensionless parameters (Pe, Sh), kinetic constants ($K$, $E_a$) and phase-dependent displacement patterns provide critical benchmarks for validating pore-scale models and predicting near-wellbore permeability impairment in geological storage of saline and hypersaline CO$_2$.

\end{abstract}

\section{Introduction}

Limiting global warming to 1.5--2\,$^\circ$C above pre-industrial levels requires rapid decarbonization of the energy sector and large-scale deployment of carbon capture and storage (CCS) technologies \citep{budinis2018assessment,nooraiepour2025norwegian,house2006permanent}. Deep saline aquifers, with their vast storage capacity and proximity to major industrial CO$_2$ sources, represent the most viable option for geological CO$_2$ sequestration at the scales needed to achieve net-zero emissions targets \citep{ringrose2021storage,michael2010geological,nooraiepour2025geological}. However, the long-term viability and economics of saline and hypersaline aquifer storage are critically dependent on maintaining stable injectivity and ensuring containment integrity throughout the operational lifetime of a storage site.

One of the most significant operational challenges to CO$_2$ injection in saline formations is salt precipitation and growth near the wellbore. This process can rapidly clog the pore space, reduce permeability, lower injection rates, and induce excess pressure buildup in the reservoir. When dry or undersaturated CO$_2$ is injected into brine-saturated reservoirs, it induces evaporation of formation water, leading to a progressive concentration of dissolved salts until supersaturation is reached and the dissolved salts crystallize within the pore space \citep{miri2016salt, dkabrowski2025surface,nooraiepour2025self}. This phenomenon, commonly termed dry\textit{ out and salting} out '' or dry out \cite{nooraiepour2018effect,miri2015new}, is most severe in the near-wellbore region where the CO$_2$ velocities are highest and the evaporation rates are maximized \citep{khosravi2024simulation,guyant2015salt,talman2020salt}. The problem is further exacerbated by capillary-driven backflow of brine to the injection point, driven by saturation gradients that develop during CO$_2$ invasion \citep{dkabrowski2025surface,ott2015salt,norouzi2022analytical,chen2024capillary}. The resulting accumulation of precipitated crystals not only reduces porosity-permeability and compromises injectivity \citep{masoudi2021pore,cui2023review,jeddizahed2016experimental,masoudi2024mineral} but can also weaken the mechanical integrity of the reservoir rock and caprock, raising concerns about long-term containment security \citep{nooraiepour2025potential,Guzina2026106404}.

Understanding and predicting CO$_2$-induced salt precipitation in saline and hypersaline aquifers requires a multi-scale multi-physics approach that integrates field observations, laboratory experiments, and numerical modeling. Reservoir-scale studies and wellbore monitoring, as in several previous studies \cite{talman2020salt,grude2014pressure}, provide direct evidence of injectivity decline and offer boundary conditions that reflect realistic geological heterogeneity and operational constraints \citep{bacci2011experimental,ringrose2021storage}. However, these observations represent integrated, spatially averaged responses that obscure the underlying pore-scale mechanisms controlling probabilistic nucleation \cite{nooraiepour2021probabilistic,nooraiepour2021Omega}, precipitation dynamics \cite{dkabrowski2025surface, nooraiepour2025self,dąbrowski2025IFT}, and growth behavior \cite{masoudi2021pore,dąbrowski2025crystallization}, which directly affect porosity-permeability deterioration \cite{masoudi2024mineral,deng2025mineral}. Core-scale flow-through experiments bridge this gap by investigating salt accumulation, permeability evolution, and breakthrough behavior under controlled pressure and temperature conditions representative of reservoir environments \citep{narayanan2023long,bacci2011experimental,akindipe2021salt}. While such experiments yield valuable empirical correlations between injection parameters and permeability impairment, they still lack the spatial and temporal resolution needed to resolve spatio-temporal dynamics of evaporation-precipitation in porous media, and the local coupling between multiphase flow and growth kinetics.

Microfluidic pore-scale experiments have emerged as a powerful complement to core-scale and field observations in porous media sciences \citep{ratanpara2025review,yang2026lab,lei2025advancing,roman2025microfluidics}, enabling direct, real-time visualization of CO$_2$-brine displacement, evaporation fronts, and crystal nucleation and growth under well-controlled thermodynamic conditions \citep{nooraiepour2018effect,dkabrowski2025surface,miri2015new,ho2020microfluidic,zhang2025pore,liu2026pore,zhang2024brine}. High-resolution imaging in transparent micromodels allows quantification of key processes that govern crystallization dynamics, including phase distributions, interfacial phenomena, and the stochastic nature of nucleation via crystallite identification. While microfluidic experiments excel at elucidating fundamental mechanisms and identifying governing dimensionless parameters, translating these insights to three-dimensional reservoir conditions requires careful dimensional analysis and phenomenological interpretation \citep{ratanpara2025review,yang2026lab,lei2025advancing,roman2025microfluidics}.

Despite advances across these scales, critical gaps remain in our quantitative understanding of how CO$_2$ phase states, temperature, and hydrodynamic transport regime jointly control crystallization kinetics in porous media. Furthermore, although permeability reduction as a function of injected pore volumes or elapsed time has been documented, the underlying crystallization kinetics—nucleation rates, growth mechanisms, and their dependence on local thermodynamic and hydrodynamic conditions remain poorly characterized. This knowledge gap hinders the development of predictive pore-scale models that can accurately simulate porosity and permeability evolution during CO$_2$ injection and provide quantitative input for upscaling to reservoir simulators.

In this study, we address these gaps by conducting high-resolution microfluidic experiments that systematically quantify the interplay between CO$_2$ phase states, temperature, and transport regime on halite crystallization kinetics during CO$_2$-driven brine evaporation. We employ borosilicate glass micromodels with rock-like pore geometries to visualize and measure brine displacement, evaporation, and crystallization under controlled conditions spanning 20--60\,$^\circ$C and 50--80\,bar, encompassing liquid, gaseous, and supercritical CO$_2$ (scCO$_2$) phases. This parameter space is representative of shallow to intermediate depth saline and hypersaline aquifers and near-wellbore conditions in actively injecting CCS projects. Halite precipitation is investigated because NaCl is the dominant dissolved salt in most formation brines and halite is the primary precipitate controlling near-wellbore permeability impairment. Flow rates are selected to achieve Péclet numbers ranging from 50 to 1440, enabling systematic evaluation of the transition from diffusion-limited to convection-dominated crystallization regimes. 

Using automated image segmentation and time-resolved analysis, we quantify key metrics including post-breakthrough brine saturation, time to first nucleation, total nucleation duration, final crystal fraction, and spatial distributions of evaporating brine and growing crystals. Dimensionless analysis based on the Péclet and Sherwood numbers characterizes mass transfer efficiency and evaporation kinetics, while the Avrami equation provides a framework for modeling crystal growth rates. Temperature-dependent kinetics are analyzed through Arrhenius plots, yielding an apparent activation energy that reflects the energetic barriers governing nucleation and growth in confined geometries. Fractal analysis of CO$_2$ invasion patterns quantifies the influence of phase state on displacement efficiency and residual saturation.

\section{Materials and Methods}

\subsection{Microfluidic Chip Design}

\begin{figure}[h!]
	\includegraphics[width=0.95\textwidth]{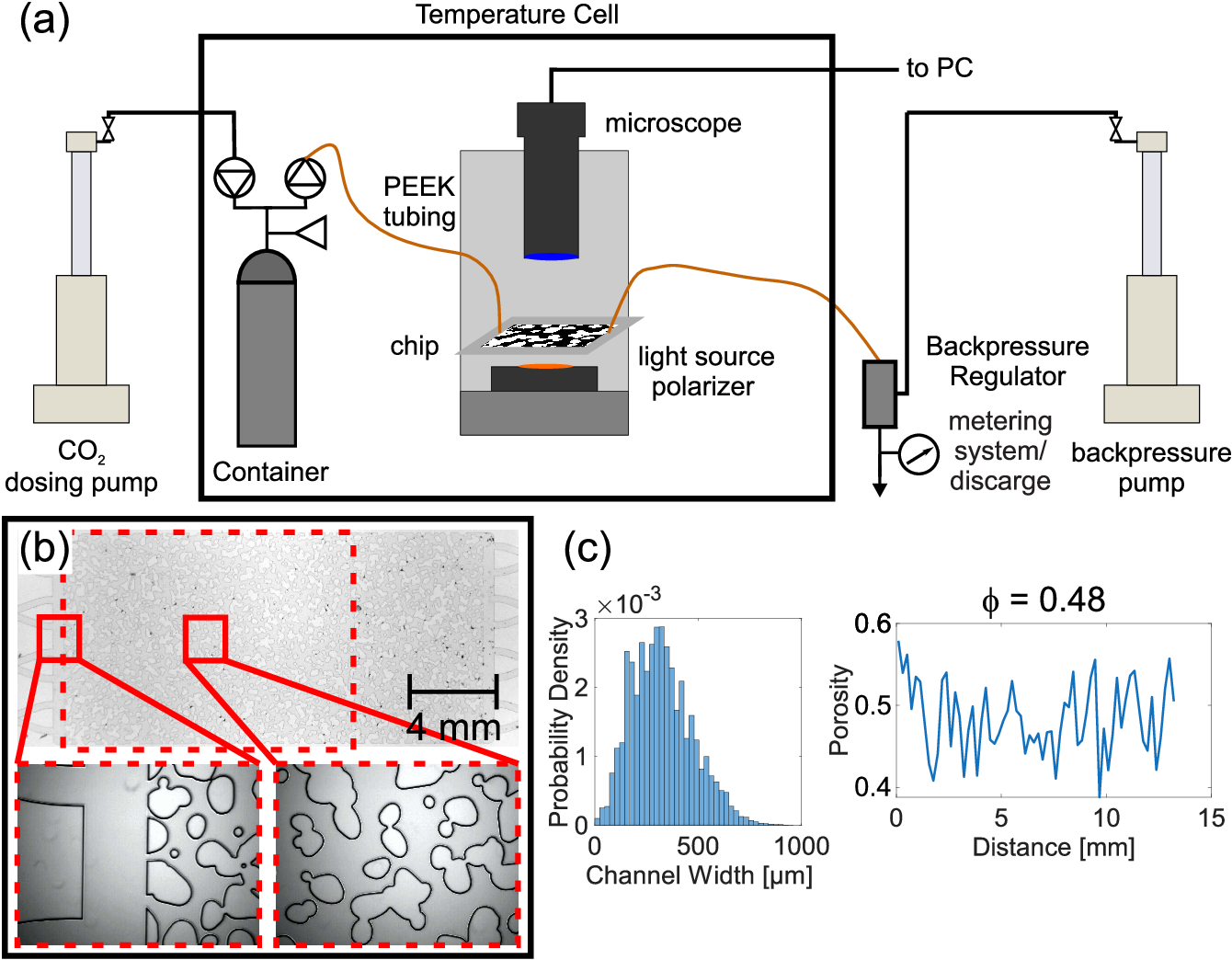}
	\caption{Experimental configuration and chip characterization. (a) Schematic diagram of the high-pressure high-temperature microfluidic system showing CO$_2$ delivery via syringe pump, temperature-controlled enclosure, microscopic imaging setup, and backpressure regulation. (b) Microscopic image of the microfluidic chip with inlet channels and porous network. The dashed rectangle indicates the field of view used for data acquisition. The inset shows a magnified view of the pore-grain structure. (c) Pore size distribution (left plot) and spatial porosity variation along the chip length (right plot).}	
	\label{fig:setup}
\end{figure}

The experiments employ a two-dimensional borosilicate glass micromodel with rock-like pore geometry fabricated using photolithography and wet etching (Micronit EOR chip) \citep{ren2013materials}. Figure~\ref{fig:setup} shows a schematic representation and microscopic image of the test setup, with an enlarged view of the inlet distribution channels and internal porous structure of the microfluidic chip provided in the inset. The chip dimensions are 20~mm (length) $\times$ 10~mm (width) $\times$ 20~$\mu$m (depth), yielding a total pore volume of 18.6~$\mu$L. The pore geometry, derived from a digitized scan of natural sandstone, exhibits a mean pore size of 240~$\pm$~135~$\mu$m with the pore size distribution shown in Figure~\ref{fig:setup}(c). The pore dimensions were deliberately upscaled relative to typical reservoir sandstones to overcome fabrication constraints and enable high-resolution optical visualization. The overall porosity is 0.48, with local spatial variations illustrated in Figure~\ref{fig:setup}(c). The manufacturer-reported permeability is 7.2~$\pm$~1.1 Darcy. Contact angle measurements for the water–air system on borosilicate glass yield 27$^\circ$, confirming strong hydrophilic behavior.

\subsection{Experimental Apparatus and Protocol}

The experimental setup is shown schematically in Figure~\ref{fig:setup}(a). The microfluidic chip was mounted inside a temperature-controlled forced convection benchtop oven (Despatch LBB series) \cite{naseryan2019relative} and connected via 1/16'' PEEK tubing to minimize dead volume and prevent unwanted brine accumulation. High-pressure CO$_2$ was supplied from a temperature-stabilized fluid transfer vessel pressurized by a syringe pump (Teledyne ISCO 100DM). Flow rate control was achieved using a downstream backpressure regulator connected to a second syringe pump (Teledyne ISCO 500D), with instantaneous flow monitoring provided by a Coriolis mass flowmeter (Burkert). 

Imaging was performed using a Dino-Lite AM73915MZT digital microscope equipped with a 5.0-megapixel CMOS sensor (2560~$\times$~1920 pixels), 10$\times$–220$\times$ optical magnification, and integrated polarized LED illumination. This configuration provided a maximum field of view of 10~$\times$~14~mm with snapshot acquisition at 1~fps throughout each experiment.

Prior to each experiment, the chip was evacuated under vacuum and then fully saturated with 5.5~mol~L$^{-1}$ halite (NaCl) brine solution prepared using deionized Milli-Q water. After saturation, the inlet was disconnected and replaced with the dry CO$_2$ supply line, eliminating residual brine in the tubing that could prematurely saturate the injected CO$_2$ or generate uncontrolled nucleation sites upstream of the chip.

The experimental procedure consisted of three stages. First, the system was pressurized to the target pressure while maintaining the outlet valve closed. Second, the outlet pressure was gradually reduced until CO$_2$ breakthrough occurred, rapidly displacing the majority of brine and leaving isolated residual pools adhered to grain surfaces by capillary forces. Third, continuous CO$_2$ flow was maintained at constant pressure and temperature while the residual brine evaporated. As evaporation proceeded, dissolved NaCl concentration in each brine pool increased until supersaturation was reached, triggering halite crystal nucleation, precipitation, and growth in porous medium. Crystal growth continued until complete dryout. Following each experiment, the chip was flushed with deionized water to dissolve precipitated salt and restore the original pore geometry.

\subsection{Image Acquisition and Segmentation}

\begin{figure}[h!]
	\includegraphics[width=0.95\textwidth]{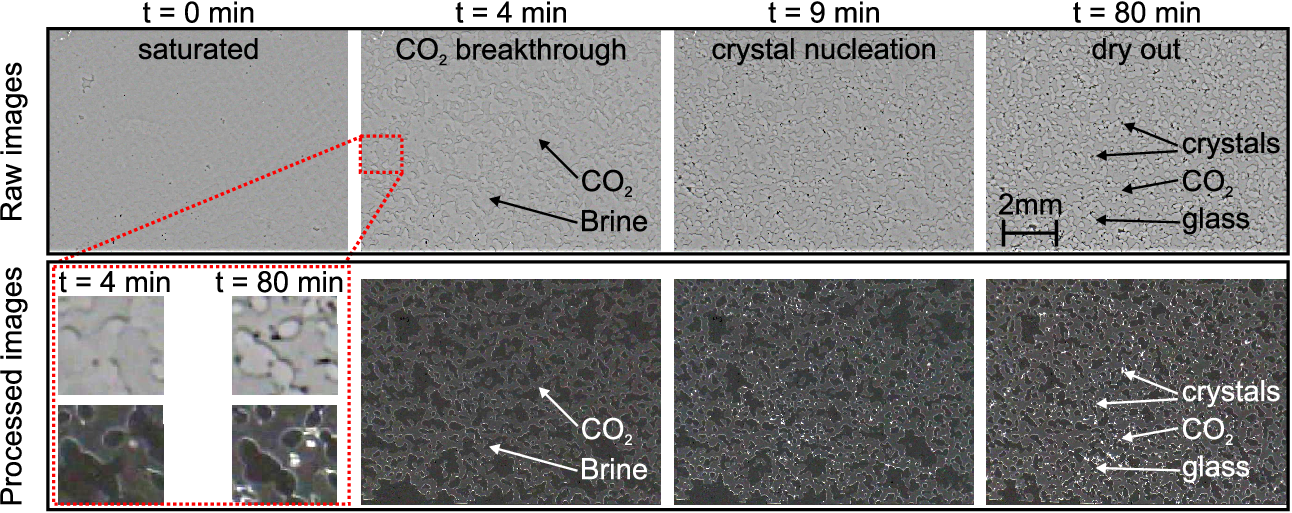}
	\caption{Image acquisition and processing workflow for a representative experiment (100~mL/min CO$_2$ flow, 80~bar, 40$^\circ$C). Upper panel: Raw microscopic images showing (left to right) initial brine saturation, post-breakthrough residual brine pools, first crystal formation, and complete dryout. The arrows identify glass grains, brine, crystals, and CO$_2$-filled pores. Lower panel: Processed images obtained by subtracting the saturated reference frame from each time-series image, enhancing contrast between phases. The inset shows a magnified view of the pore-scale crystal distribution across two time stages. Note the minimal contrast between brine and glass in raw images, necessitating image subtraction for accurate segmentation.}	
	\label{fig:RawandProcImages}
\end{figure}

Figure~\ref{fig:RawandProcImages} illustrates the image processing workflow. The upper panel shows raw microscopic images at four key stages: (i) initial brine saturation, (ii) post-breakthrough with residual brine pools, (iii) onset of crystal formation, and (iv) complete dryout. In the saturated state, brine and glass are nearly indistinguishable due to similar refractive indices (1.33 and 1.47, respectively), rendering the pore network invisible. After the CO$_2$ breakthrough, CO$_2$-filled pore space appears as darker regions while brine-filled regions remain bright. Upon crystallization, halite crystals appear as high-contrast black spots. After complete evaporation, only glass grains, crystalline salt, and CO$_2$-filled pores remain, each marked with separate arrows in Figure~\ref{fig:RawandProcImages}.

To enhance phase contrast, each frame was processed by subtracting the initial fully saturated reference image. In the resulting processed images (lower panel of Figure~\ref{fig:RawandProcImages}), brine and glass exhibit low normalized intensity ($I \approx 0$, black), crystals exhibit high intensity ($I \approx 1$, white), and CO$_2$-filled pores show intermediate values. 

Phase segmentation was performed using a custom MATLAB image processing pipeline based on intensity thresholding. First, a glass mask was generated from the dry-chip image by identifying pixels with intensity below a threshold. Crystal regions were then segmented from each time-series frame as pixels exceeding an upper threshold. Finally, brine regions were identified as low-intensity pixels not classified as glass. From these binary masks, we computed porosity $\phi = A_p / A$, brine saturation $S_w = A_w / A_p$, and crystal fraction $X_c = A_c / A_p$, where $A$ is the total chip area, $A_p$ is the pore area, $A_w$ is the brine-occupied area, and $A_c$ is the crystal-occupied area. Temporal evolution of $S_w(t)$ and $X_c(t)$ constitutes the primary quantitative output of this analysis.

\subsection{Experimental Conditions and Dimensionless Parameters}

Experiments were conducted across a matrix of thermodynamic conditions designed to span liquid, gaseous, and supercritical CO$_2$ phases, as summarized in Table~\ref{tab:ExpCond}. Pressure ranged from 50 to 80~bar, temperature from 20 to 60$^\circ$C, and volumetric flow rate (measured at ambient conditions) was either 100 or 1000~mL/min. Each condition was replicated three times to assess reproducibility. This parameter space enables systematic investigation of how CO$_2$ phase properties—density, viscosity, and diffusivity—influence displacement efficiency, evaporation kinetics, and crystallization dynamics.

Although flow rates were fixed at ambient pressure, the in-situ interstitial velocity $u_m$ varies with thermodynamic conditions due to density changes:
\begin{equation}
u_m = \frac{V_{\text{in-situ}}}{h \cdot b \cdot \phi}
\end{equation}
where $h$ is the chip depth (20~$\mu$m), $b$ is the chip width (10~mm), $\phi$ is porosity (0.48), and $V_{\text{in-situ}}$ is the volumetric flow rate corrected for in-situ density. The resulting interstitial velocities range from 0.1 to 0.7~m/s, which correspond to near-wellbore conditions in active CCS projects. For example, assuming an injection rate of 5~Mt CO$_2$/year (planned for Phase 2 of the Northern Lights CCS project \citep{acuna2024northern}) distributed over a 20-m perforation interval, these velocities are representative of conditions within approximately 0.1~m of the wellbore.

Flow characteristics are quantified using the Reynolds number:
\begin{equation}
\text{Re} = \frac{\rho u_m D_h}{\mu}
\end{equation}
where $D_h = 2hb/(h+b)$ is the hydraulic diameter, $\rho$ is the CO$_2$ density, and $\mu$ is the dynamic viscosity. As shown in Table~\ref{tab:ExpCond}, Re ranges from 17 to 695, confirming that all experiments operate in the laminar flow regime where viscous forces dominate inertial effects.

Mass transport is characterized by the Péclet number, which quantifies the relative importance of convective versus diffusive transport:
\begin{equation}
\text{Pe} = \frac{u_m D_h}{D} = \text{Re} \cdot \text{Sc}, 
\quad 
\text{Sc} = \frac{\mu}{\rho D}
\end{equation}
where $D$ is the molecular diffusion coefficient of water vapor in CO$_2$ and Sc is the Schmidt number. Table~\ref{tab:ExpCond} shows that Pe ranges from 50 to 1440, indicating that convective transport dominates diffusion across all conditions. Notably, Pe varies most strongly with flow rate, suggesting that flow velocity is the primary control on evaporation and crystallization kinetics. Secondary effects arise from phase-dependent variations in density and viscosity, which influence both Re and Sc.

\begin{table}[h!]
    \centering
	\caption{Experimental conditions and dimensionless flow parameters for liquid, gaseous, and supercritical CO$_2$ reactive flow experiments. Reynolds (Re) and Péclet (Pe) numbers characterize flow regime and mass transport, respectively. Viscosity and density values are calculated from the NIST REFPROP database \cite{huber2022nist,lemmon2018nist}.}
	\footnotesize
	\begin{tabular}{c c c c c c c c}
		\toprule
		Pressure & Temperature & Viscosity & Density & Flow Rate & Re & Pe & Phase \\
		(bar) & ($^\circ$C) & ($\times 10^{-5}$ Pa·s) & (kg/m$^3$) & (mL/min) & & & \\
		\midrule
		\multicolumn{8}{c}{\textit{Flow Rate = 100 mL/min}} \\
		50  & 20 & 1.663 & 140.6  & 100 & 75  & 108  & Gas \\
		60  & 20 & 6.972 & 782.7  & 100 & 18  & 56 & Liquid \\
		60  & 40 & 1.780 & 149.3  & 100 & 69  & 86 & Gas \\
		60  & 60 & 1.818 & 124.9  & 100 & 68 & 69  & Gas \\
		65  & 20 & 7.213 & 795.7  & 100 & 17 & 50 & Liquid \\
		80  & 40 & 2.230 & 277.9  & 100 & 56 & 144 & Supercritical \\
		80  & 60 & 1.995 & 191.6  & 100 & 62 & 78  & Supercritical \\
		\midrule
		\multicolumn{8}{c}{\textit{Flow Rate = 1000 mL/min}} \\
		60  & 20 & 6.972 & 782.7  & 1000 & 177 & 555 & Liquid \\
		60  & 40 & 1.780 & 149.3  & 1000 & 695 & 863 & Gas \\
		60  & 60 & 1.818 & 124.9  & 1000 & 680 & 698 & Gas \\
		80  & 20 & 7.833 & 827.7  & 1000 & 158 & 393 & Liquid \\
		80  & 40 & 2.230 & 277.9  & 1000 & 554 & 1440 & Supercritical \\
		80  & 60 & 1.995 & 191.6  & 1000 & 620 & 780 & Supercritical \\
		\bottomrule
	\end{tabular}
	\label{tab:ExpCond}
\end{table}

\subsection{Analytical Framework: Evaporation and Crystallization Kinetics}

\subsubsection{Evaporation Modeling and Sherwood Number}

Brine evaporation is quantified through the mass transfer coefficient $k_c$, which relates the evaporative flux to the concentration driving force. This is expressed via the dimensionless Sherwood number:
\begin{equation}
\text{Sh} = \frac{k_c D_h}{D}, 
\quad
k_c = \frac{J_{\text{avg}}}{c_{\text{sat}} - c_\infty}
\label{eq:Scherwood}
\end{equation}
where $J_{\text{avg}}$ is the mean evaporation flux (mass per unit area per unit time), $c_{\text{sat}}$ is the water vapor concentration at the brine–CO$_2$ interface (assumed to be saturated), and $c_\infty$ is the bulk concentration in the flowing CO$_2$ stream (assumed negligible for dry injection). The Sherwood number quantifies the enhancement of mass transfer due to convection relative to pure diffusion; Sh $\approx$ 1 indicates diffusion-limited transport, while Sh $\gg$ 1 indicates convection-dominated transport.

For thin liquid films evaporating under steady convective flow, the temporal evolution of brine-covered area follows an exponential decay:
\begin{equation}
A(t) = A_0 \exp\left(- \frac{J_{\text{avg}}}{h \rho_w} t \right)
\label{eq:evaporationModel}
\end{equation}
where $A_0$ is the initial brine-covered area immediately after CO$_2$ breakthrough, $h$ is the chip depth, and $\rho_w$ is the brine density. By fitting this expression to the measured $S_w(t)$ data, we extract $J_{\text{avg}}$, from which Sh is computed using Equation~\ref{eq:Scherwood}.

\subsubsection{Fractal Analysis of CO$_2$ Invasion Patterns}

The spatial complexity of the CO$_2$ invasion pattern after breakthrough is characterized by the fractal dimension $D$, which quantifies how completely the invading fluid fills the available pore space. Higher fractal dimensions indicate more uniform, space-filling invasion and are associated with lower residual brine saturation. The fractal dimension is computed using the box-counting method:
\begin{equation}
D = \lim_{\epsilon \to 0} \frac{\log N(\epsilon)}{\log (1/\epsilon)}
\label{eq:fractal}
\end{equation}
where $N(\epsilon)$ is the number of boxes of size $\epsilon$ required to cover the CO$_2$-occupied region. In practice, $D$ is determined from the slope of a log–log plot of $N(\epsilon)$ versus $1/\epsilon$ over a range of box sizes. The fractal dimension provides insight into how CO$_2$ phase state and flow conditions influence displacement efficiency and pore-scale connectivity.

\subsubsection{Crystal Growth Kinetics: Avrami Model}

Crystallization kinetics are modeled using the Avrami equation, which describes the time evolution of the crystallized fraction:
\begin{equation}
X_c(t) = 1 - \exp\left(-K t^n\right)
\label{eq:avrami}
\end{equation}
where $X_c(t)$ is the volume fraction of crystallized salt at time $t$, $K$ is the Avrami rate constant (units: s$^{-n}$), and $n$ is the Avrami exponent. The exponent $n$ reflects the dimensionality of crystal growth and the nucleation mechanism: $n \approx 1$ indicates one-dimensional (linear) growth, $n \approx 2$ indicates surface-controlled growth in two dimensions, and $n \approx 3$ indicates three-dimensional volumetric growth. Given the quasi-two-dimensional geometry of the microfluidic chip, we expect $n \approx 2$.

The Avrami constant $K$ encapsulates both nucleation and growth rates and is expected to depend on temperature and transport conditions. By fitting Equation~\ref{eq:avrami} to the measured $X_c(t)$ curves, we extract $K$ and $n$ for each experimental condition, enabling systematic comparison across CO$_2$ phases, temperatures, and Péclet numbers.

\subsubsection{Temperature Dependence and Activation Energy}

The temperature dependence of the Avrami rate constant is described by the Arrhenius equation:
\begin{equation}
K(T) = K_0 \exp\left(-\frac{E_a}{R T}\right)
\end{equation}
where $K_0$ is the pre-exponential factor, $E_a$ is the apparent activation energy (J/mol), $R$ is the universal gas constant (8.314~J/(mol·K)), and $T$ is the absolute temperature (K). Taking the natural logarithm yields a linearized form:
\begin{equation}
\ln K = \ln K_0 - \frac{E_a}{R} \cdot \frac{1}{T}
\label{eq:activationLinear}
\end{equation}
from which $E_a$ and $K_0$ are determined by linear regression of $\ln K$ versus $1/T$ (Arrhenius plot). The slope of this plot is $-E_a/R$ and the intercept is $\ln K_0$.

The activation energy is the minimum energy barrier for nucleation and growth. A high $E_a$ indicates strong temperature sensitivity, meaning that crystallization accelerates rapidly with increasing temperature. Conversely, a low $E_a$ suggests that the process is less thermally activated and may be controlled by transport limitations rather than kinetic barriers. By extracting $E_a$ from our experiments, we can assess whether crystallization in microfluidic geometries is controlled by interfacial kinetics (high $E_a$) or by convective/diffusive transport (low $E_a$). These insights are critical for predicting crystallization behavior under variable reservoir thermal conditions and for optimizing injection strategies to minimize near-wellbore permeability impairment.

\section{Results and Discussion}

\subsection{CO$_2$ Breakthrough and Displacement Efficiency}

\begin{figure}[ht]
	\includegraphics[width=0.95\textwidth]{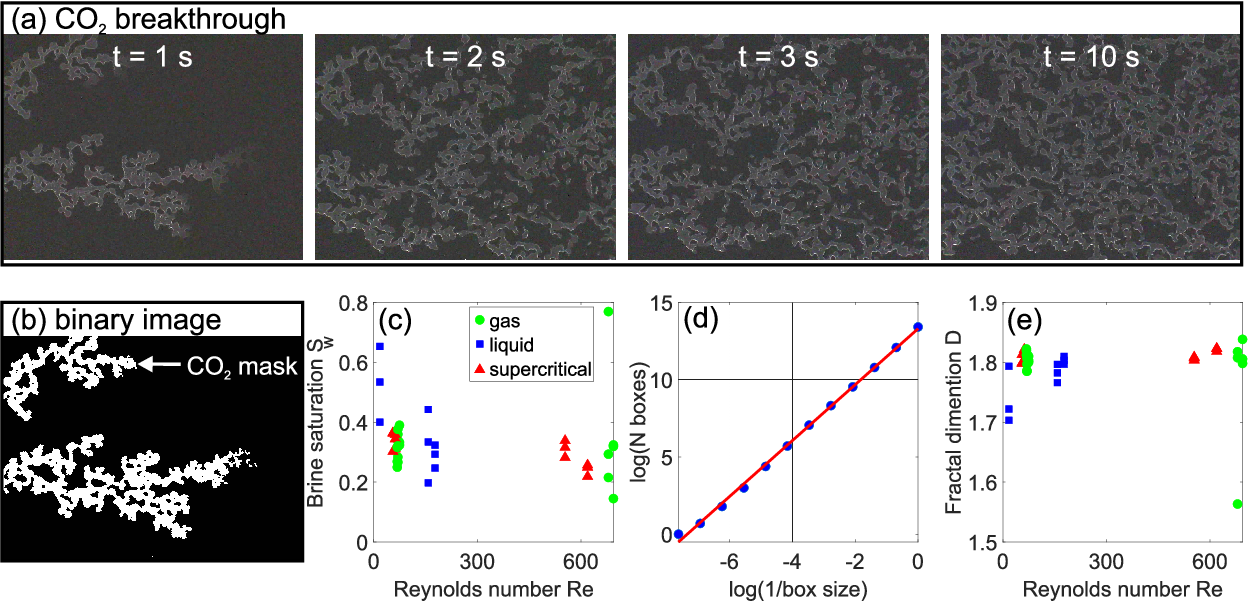}
	\caption{CO$_2$ breakthrough dynamics and displacement efficiency analysis. (a) Time-lapse contrast images showing CO$_2$ invasion at 100~mL/min, 80~bar, 40$^\circ$C. It shows the tree-like branching structure and progressive brine isolation. (b) Binary mask of CO$_2$-occupied pore space extracted from panel (a). (c) Post-breakthrough brine saturation versus Reynolds number for gaseous (green), liquid (blue), and supercritical (red) CO$_2$ phase states. Supercritical conditions yield the lowest mean saturation and the smallest variability. (d) Representative box-counting analysis for fractal dimension determination. The slope of the log--log plot yields $D$~=~1.78 for this example. (e) Fractal dimension of CO$_2$ invasion patterns versus Reynolds number. Higher $D$ correlates with more connected flow networks and lower residual saturation, with supercritical CO$_2$ approaching the dry-chip limit ($D$~=~1.85).}	
	\label{fig:CO2Breakthrough}
\end{figure}

CO$_2$ breakthrough rapidly desaturates the initially brine-saturated chip within seconds, displacing the mobile brine fraction and leaving behind isolated residual pools trapped by capillary forces at grain surfaces. The post-breakthrough brine saturation ($S_w$) and spatial distribution of these residual pools critically determine subsequent evaporation kinetics and crystallization patterns. Figure~\ref{fig:CO2Breakthrough}(a) shows the temporal evolution of a typical breakthrough event for 100~mL/min CO$_2$ flow at 80~bar and 40$^\circ$C. The CO$_2$ invasion pattern exhibits a characteristic branched structure rooted in the inlet channels and extending toward the outlet, progressively isolating brine into disconnected pools. Binary segmentation of the CO$_2$ phase (Figure~\ref{fig:CO2Breakthrough}(b)) enables quantification of $S_w$ as a function of time, with breakthrough completion defined as the point where rapid desaturation ceases.

Figure~\ref{fig:CO2Breakthrough}(c) presents post-breakthrough saturation as a function of Reynolds number for all experimental conditions, with different symbols denoting liquid, gaseous, and supercritical CO$_2$ (scCO$_2$) phase states. Several key trends emerge. First, substantial variability is observed even at fixed Reynolds numbers, reflecting the stochastic nature of capillary trapping in heterogeneous pore networks. For liquid CO$_2$ at Re~$\approx$~18, $S_w$ ranges from 0.40 to 0.62 (37\% variation), whereas at higher Reynolds numbers (Re~$\approx$~160--180), saturation decreases to 0.20--0.40 but with comparable scatter (25--45\% variation). Gaseous CO$_2$ achieves lower residual saturations, ranging from 0.24--0.39 at Re~$\approx$~68--75 to 0.14--0.32 at Re~$\approx$~680--695 (excluding one outlier at $S_w$~=~0.79). 

Notably, scCO$_2$ exhibits both the lowest mean saturation and the smallest variability, with $S_w$~=~0.30--0.36 at Re~$\approx$~56--62 (20\% variation) and $S_w$~=~0.22--0.34 at higher Reynolds numbers (15--25\% variation). This reduced scatter suggests that supercritical conditions promote more reproducible displacement, likely due to favorable viscosity ratios and interfacial properties that minimize capillary trapping heterogeneity. While higher Reynolds numbers generally correlate with lower $S_w$, the most dramatic reduction occurs in the transition from low-Re liquid conditions (Re~$\approx$~18) to higher-Re flows, with diminishing returns at Re~$>$~100. This trend aligns with the capillary number dependence of residual saturation in drainage processes, where increased viscous forces progressively overcome capillary retention up to a threshold beyond which further increases yield marginal improvements.

To characterize the spatial connectivity and complexity of the CO$_2$ invasion patterns, we computed the fractal dimension ($D$) of post-breakthrough CO$_2$ masks using the box-counting method (Equation~\ref{eq:fractal}). Figure~\ref{fig:CO2Breakthrough}(d) shows a representative log--log plot of $N(\epsilon)$ versus $1/\epsilon$, where the slope of the linear fit yields $D$. The dry chip exhibits $D$~=~1.85, indicating a highly connected pore network. After brine saturation and subsequent CO$_2$ invasion, $D$ decreases due to partial channel blockage by residual brine, with the extent of reduction correlating inversely with displacement efficiency.

Figure~\ref{fig:CO2Breakthrough}(e) presents fractal dimensions across all experiments. Liquid CO$_2$ at Re~$\approx$~18 yields the lowest values ($D$~=~1.70--1.80), consistent with poor displacement and high residual saturation. At higher Reynolds numbers (Re~$\approx$~160--180), liquid-phase fractal dimensions increase to 1.76--1.80, reflecting improved but still incomplete sweep. Gaseous CO$_2$ achieves higher fractal dimensions ($D$~=~1.75--1.82 at Re~$\approx$~68--75; $D$~=~1.79--1.84 at Re~$\approx$~680--695), indicating more pervasive invasion and better-connected flow pathways. The scCO$_2$ consistently yields the highest fractal dimensions ($D$~=~1.79--1.82), approaching the dry-chip value and confirming superior displacement efficiency.

The correlation between fractal dimension and residual saturation is physically intuitive: higher $D$ values correspond to more space-filling, uniform invasion patterns that minimize isolated brine clusters, thereby reducing $S_w$. The phase dependence reflects the combined influence of viscosity ratio, interfacial tension, and density contrast on displacement mechanisms and sweep efficiency in porous media. ScCO$_2$ achieves favorable mobility ratios that suppress viscous fingering and promote more stable displacement fronts.

\subsection{Phase-Dependent Evaporation and Crystallization Dynamics}

Following the breakthrough, residual brine evaporates progressively under continuous CO$_2$ flow, driving local salinity increases until supersaturation is reached, triggering halite crystal nucleation. The nucleation events are inherently probabilistic, leading to heterogeneous crystallization patterns even under nominally identical thermodynamic conditions. Because brine pools are spatially isolated and vary in size, this results in stochastic behavior in both the spatial and temporal domains. CO$_2$ phase state and temperature strongly influence evaporation rates, nucleation kinetics, and final crystal distributions.

\begin{figure}[h!]
	\includegraphics[width=0.95\textwidth]{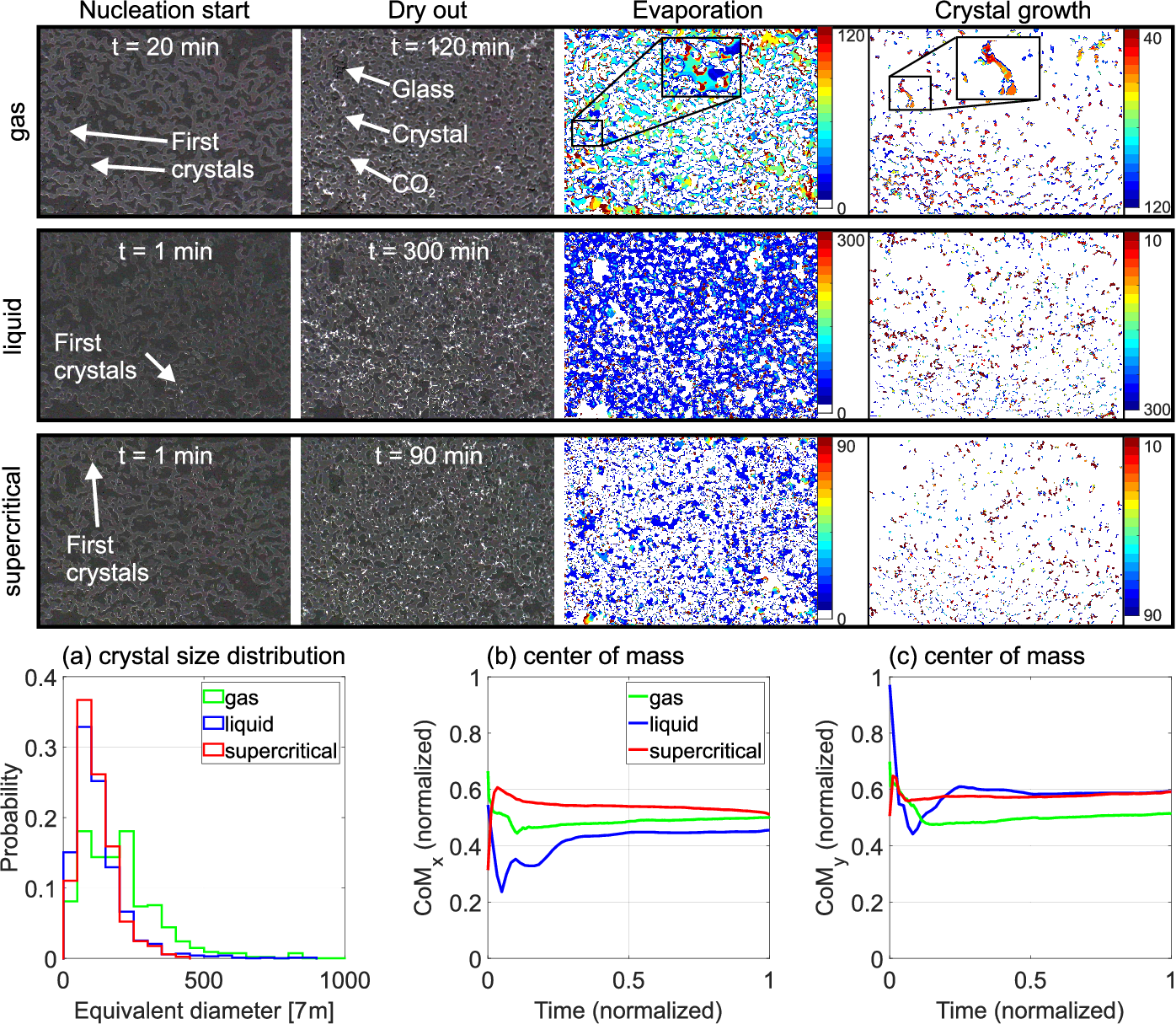}
	\caption{Comparative evaporation and halite precipitation dynamics for gaseous (50~bar, 20$^\circ$C), liquid (65~bar, 20$^\circ$C), and supercritical (80~bar, 40$^\circ$C) CO$_2$ at 1000~mL/min. First column: First nucleation events (white arrows mark crystals). Note earlier nucleation for liquid/supercritical phases (1~min) versus gas (20~min). Second column: Final crystal distributions after complete dryout. Third column: Spatiotemporal evaporation maps (blue: early; red: late). The supercritical and liquid phases show predominantly early evaporation, whereas the gas phase exhibits more sustained evaporation. Fourth column: Crystal growth maps (red: early; blue: late). Insets show single-pore crystal evolution. (a) Equivalent diameter distributions for final crystals. Gaseous CO$_2$ produces larger, more heterogeneous crystals. (b,c) Normalized center-of-mass trajectories in flow-parallel and flow-perpendicular directions. Center of mass convergence to 0.5 indicates spatially uniform final distributions despite stochastic nucleation.}	
	\label{fig:TimeEvolutionThreePhases}
\end{figure}

Figure~\ref{fig:TimeEvolutionThreePhases} compares evaporation and crystallization for gaseous (50~bar, 20$^\circ$C), liquid (65~bar, 20$^\circ$C), and supercritical (80~bar, 40$^\circ$C) CO$_2$ at 1000~mL/min. The first column shows microscopic images at the onset of nucleation, with crystallite formation identified (the first crystals are marked with white arrows). For gaseous CO$_2$, nucleation occurs 20~min post-breakthrough at two spatially separated sites. In contrast, both liquid and supercritical phases nucleate within 1~min, indicating much faster supersaturation development. Importantly, nucleation sites differ across replicates under identical conditions, confirming the stochastic nature of heterogeneous nucleation in confined porous geometries. The lower image contrast for liquid CO$_2$ reflects its higher refractive index relative to gas and supercritical phases.

The second column shows final states after complete dryout, with glass, crystals, and CO$_2$-filled pores identified. Total evaporation times vary significantly: 90~min for scCO$_2$, 220~min for liquid CO$_2$, and 300~min for gaseous CO$_2$. The faster dryout under supercritical conditions highlights the dominant role of enhanced mass transfer, driven by high diffusivity and favorable viscosity in this regime.

The third column presents spatiotemporal evaporation maps, where colors encode the timing of local dryout (blue: early; red: late). Regions that evaporate rapidly in the first minute (prior to the color-mapping threshold) are excluded for clarity. For all phases, evaporation initially proceeds uniformly but slows as brine retreats into narrow ridges and concave pore corners where capillary retention is strongest. Gaseous CO$_2$ exhibits relatively uniform evaporation throughout the process, with significant fractions of blue, yellow, and red regions indicating sustained evaporation over time. Liquid CO$_2$ shows predominantly blue regions, indicating that most evaporation occurs early, with smaller late-stage pools persisting in isolated geometries. The scCO$_2$ displays a pattern similar to liquid, but with even more rapid early-stage evaporation, reflected in the dominance of blue and green regions and minimal late-stage brine residues.

The fourth column shows crystal growth maps, where red denotes early-forming crystals, and blue indicates late-stage precipitation. Insets illustrate the evolution of individual crystal patches within adjacent pores. The majority of crystallization occurs during the early evaporation phase, coinciding with the period of fastest brine concentration increase. However, crystal growth continues throughout the experiment, with large crystals growing incrementally and new nuclei forming sporadically. Final crystal fractions are 0.067 (gas), 0.068 (liquid), and 0.046 (supercritical). The lower value for scCO$_2$ may reflect more efficient brine removal during breakthrough, leaving less salt available as dissolved solutes for precipitation despite faster evaporation.

Figure~\ref{fig:TimeEvolutionThreePhases}(a) presents equivalent diameter distributions for the final crystal populations, calculated as $D_{\text{eq}} = 2\sqrt{A/\pi}$ where $A$ is the projected crystal area. Gaseous CO$_2$ yields larger crystals with a broader distribution (mean $\pm$ std: 200~$\pm$~150~$\mu$m), whereas liquid and supercritical phases produce smaller, more uniform crystals (130~$\pm$~90~$\mu$m and 120~$\pm$~70~$\mu$m, respectively). The twofold difference in mean crystal size between gas and liquid/supercritical phases occurs despite similar final crystal fractions in the gas and liquid cases, indicating that slower evaporation promotes growth of fewer, larger crystals, while faster evaporation favors nucleation of many smaller crystals, consistent with classical nucleation theory where high supersaturation rates increase nucleation frequency.

To assess spatial uniformity of crystallization, we computed the time-dependent center of mass (CoM) of the crystal distribution in both flow-parallel ($x$) and flow-perpendicular ($y$) directions:
\begin{equation}
\text{CoM}_x(t) = \frac{\sum_{x,y} x \, M_t(x,y)}{\sum_{x,y} M_t(x,y)}, \quad \text{CoM}_x^*(t) = \frac{\text{CoM}_x(t)}{N_x}
\end{equation}
where $M_t(x,y)$ is the binary crystal mask at time $t$, and normalization by image dimensions ($N_x$, $N_y$) yields dimensionless coordinates ranging from 0 to 1. Figure~\ref{fig:TimeEvolutionThreePhases}(b,c) shows CoM normalized temporal evolution for the three phases. 

In the flow direction ($x$), initial CoM positions exhibit phase-dependent biases: gas (0.54, slight outlet bias), liquid (0.31, significant inlet bias), and supercritical (0.67, strong outlet bias). As crystallization proceeds, CoM values converge asymptotically toward 0.5, with final values of 0.52 (gas), 0.50 (liquid), and 0.45 (supercritical), indicating spatially uniform final distributions with no systematic inlet-to-outlet progression. The absence of directional crystallization reflects the small chip dimensions, where evaporation timescales are comparable across all locations. In reservoir-scale systems, by contrast, crystallization is expected to propagate radially from injection wells due to sustained evaporation fronts and brine capillary backflow. This highlights a limitation of microfluidic studies for capturing large-scale reactive transport phenomena, though the mechanistic insights into nucleation and growth remain valid.

In the transverse direction ($y$), stronger initial biases are observed for gas (0.97) and liquid (0.69), while supercritical shows no bias (0.50). Final CoM positions are 0.59 (gas), 0.59 (supercritical), and 0.51 (supercritical), suggesting mild asymmetries that may reflect local porosity variations in the utilized chip or imaging field-of-view effects rather than systematic physical mechanisms.

\begin{figure}[ht]
	\includegraphics[width=0.95\textwidth]{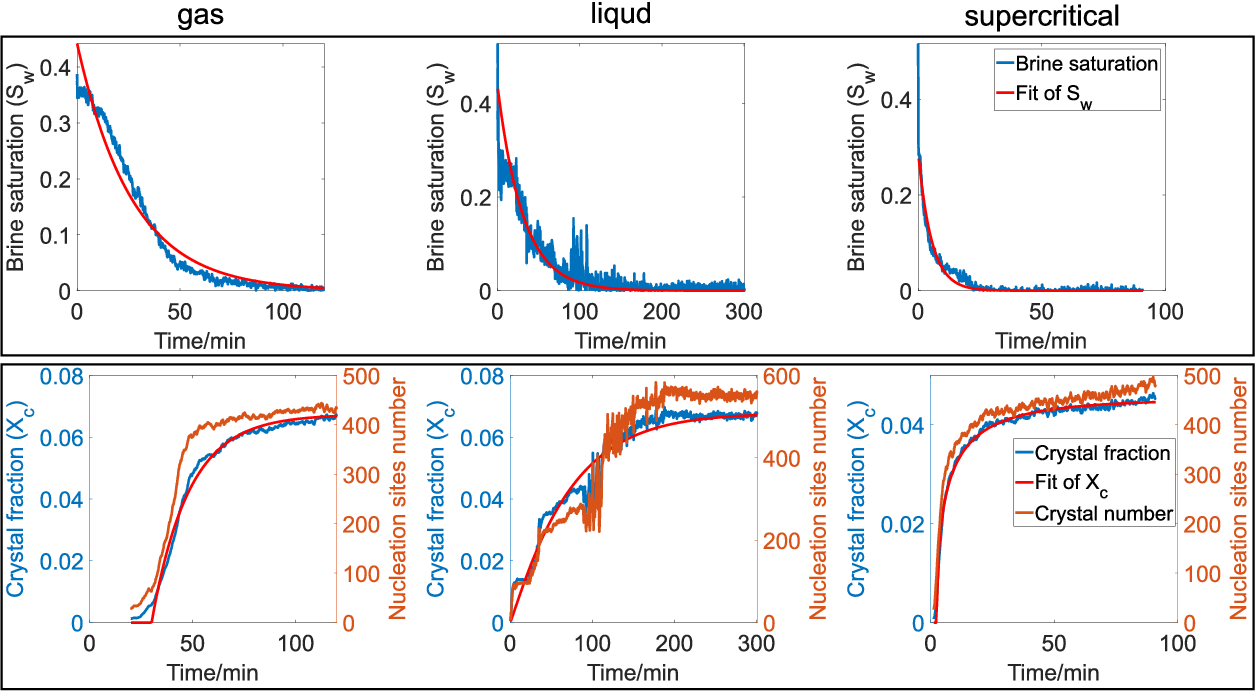}
	\caption{Temporal evolution of brine saturation and crystallization for gaseous, liquid, and supercritical CO$_2$ (conditions as in Figure~\ref{fig:TimeEvolutionThreePhases}). Upper panel: Brine saturation $S_w(t)$ (blue) and exponential decay model fits (Equation~\ref{eq:evaporationModel}, red). Gaseous CO$_2$ exhibits more sustained evaporation, while liquid and supercritical phases show rapid early-stage drying. Lower panel: Crystal fraction $X_c(t)$ (blue) with Avrami model fits ($n$~=~3, Equation~\ref{eq:avrami}, red) and detected crystal count (orange, right axis). Early-stage crystallization dominates for all phases. Non-monotonic crystal count reflects coalescence and detection limitations.}	
	\label{fig:profilesEvaporationandcrystalisation}
\end{figure}

Figure~\ref{fig:profilesEvaporationandcrystalisation} quantifies the temporal evolution of brine saturation and halite crystal formation for the three CO$_2$ phase states shown in Figure~\ref{fig:TimeEvolutionThreePhases}. The upper panel shows $S_w(t)$ (blue) alongside exponential decay model fits (Equation~\ref{eq:evaporationModel}, red). The steep initial drop reflects CO$_2$ breakthrough, followed by gradual evaporation that decelerates as brine retreats into capillary-trapped configurations. Gaseous CO$_2$ exhibits more uniform evaporation over time compared to liquid and supercritical phases, where early-stage evaporation dominates. Initial post-breakthrough saturations are 0.35 (gas), 0.27 (liquid), and 0.30 (supercritical), consistent with the phase-dependent displacement efficiencies discussed above.

From the exponential fits, we extract mean evaporation fluxes ($J_{\text{avg}}$) and compute Sherwood numbers (Equation~\ref{eq:Scherwood}), which quantify convective mass transfer enhancement relative to pure diffusion. These results are analyzed systematically as a function of Péclet number in the next sections.

The lower panel shows crystal formation fraction over elapsed time $X_c(t)$ (blue) with Avrami model fits (Equation~\ref{eq:avrami}, red, using $n$~=~3) and the number of detected crystals (orange). Crystal fraction increases most rapidly during early evaporation, particularly for scCO$_2$, while liquid CO$_2$ shows steadier growth and gaseous CO$_2$ exhibits intermediate behavior. The evolution of crystal count reveals non-monotonic trends due to crystal coalescence and detection limitations for sub-pixel crystals. Final crystal numbers are lowest for gas, intermediate for supercritical, and highest for liquid, despite similar $X_c$ values for gas and liquid. This decoupling of crystal number from crystal fraction reflects the trade-off between nucleation frequency and individual crystal size: slower evaporation (gas) favors the growth of fewer large crystals, while faster evaporation (liquid, supercritical) promotes the formation of many small crystals. From the Avrami fits, we extract rate constants $K$ that quantify the overall crystallization kinetics, which are analyzed in the following sections.

\subsection{Temperature Effects on Nucleation and Evaporation Kinetics}

\begin{figure}[ht]
	\includegraphics[width=0.95\textwidth]{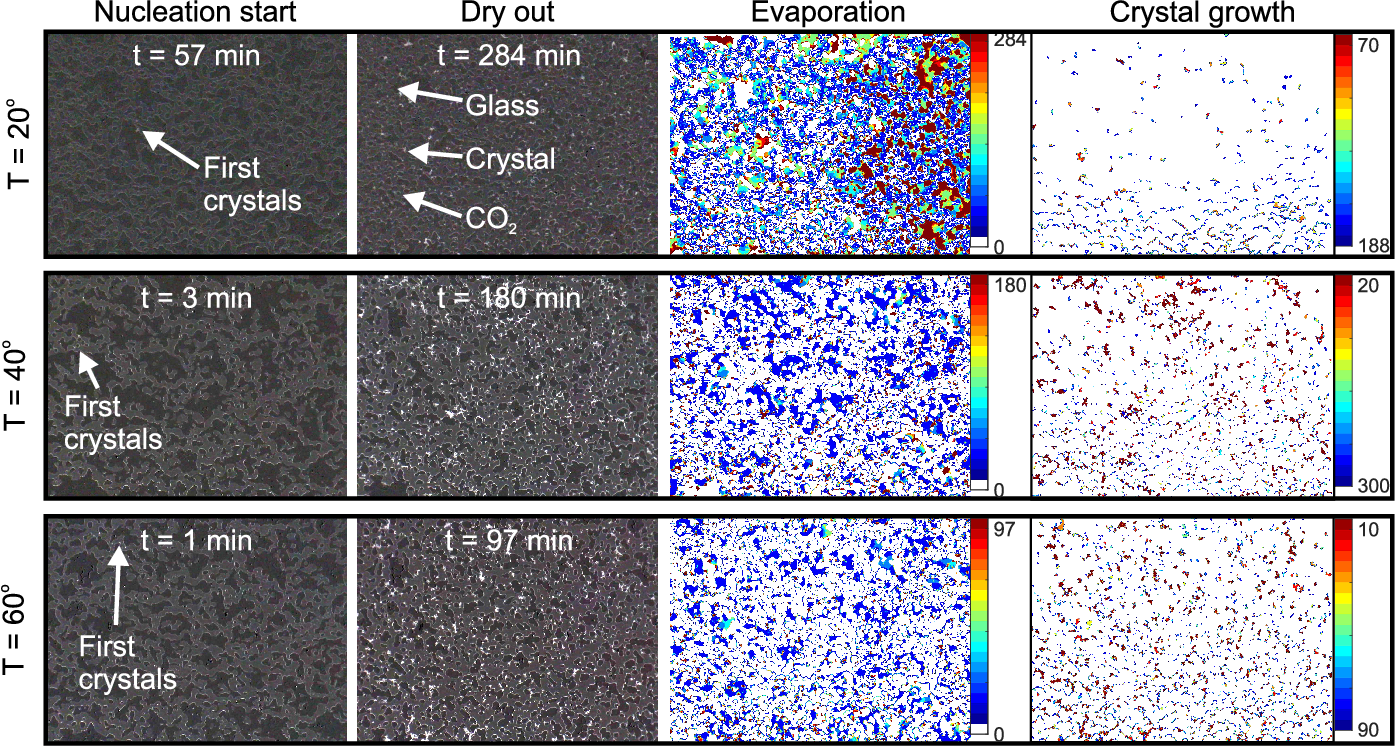}
	\caption{Temperature dependence of crystallization at fixed pressure (60~bar) and flow rate (100~mL/min). Rows correspond to 20$^\circ$C (liquid CO$_2$), 40$^\circ$C (gaseous CO$_2$), and 60$^\circ$C (gaseous CO$_2$). First column: Nucleation onset (white arrows mark first crystals). Nucleation time decreases from 57~min (20$^\circ$C) to 3~min (40$^\circ$C) to 1~min (60$^\circ$C). Second column: Final crystal distributions after complete dryout (glass, crystals, CO$_2$ marked). Total evaporation times: 284~min (20$^\circ$C), 180~min (40$^\circ$C), 97~min (60$^\circ$C). Third column: Spatiotemporal evaporation maps (blue: early; red: late). Higher temperatures yield faster, more uniform evaporation. Fourth column: Crystal growth maps (red: early; blue: late). Elevated temperatures produce denser, more heterogeneous crystal distributions. Insets show pore-scale crystal evolution.}	
	\label{fig:TemperatureImpact}
\end{figure}

To isolate the effect of temperature on crystallization and growth dynamics, Figure~\ref{fig:TemperatureImpact} compares three experiments at fixed pressure (60~bar) and flow rate (100~mL/min) but varying temperature: 20$^\circ$C (liquid CO$_2$), 40$^\circ$C (gaseous CO$_2$), and 60$^\circ$C (gaseous CO$_2$). This series spans a phase transition and a substantial temperature gradient while maintaining similar hydrodynamic conditions, enabling direct assessment of thermal activation effects.

Post-breakthrough brine saturations decrease systematically with temperature: 0.65 at 20$^\circ$C, 0.32 at 40$^\circ$C, and 0.25 at 60$^\circ$C. This trend reflects temperature-dependent changes in interfacial tension, viscosity ratio, and CO$_2$ density that collectively enhance drainage efficiency at elevated temperatures. Correspondingly, final crystal fractions increase dramatically from 0.008 at 20$^\circ$C to 0.080 at 40$^\circ$C and 0.074 at 60$^\circ$C, demonstrating that gas-phase conditions promote far greater salt precipitation than liquid-phase conditions under otherwise similar flow rates.

The first column of Figure~\ref{fig:TemperatureImpact} shows nucleation onset (via crystallite formation as a proxy for stable nucleation events), with first crystals marked by white arrows. At 20$^\circ$C, nucleation occurs 57~min post-breakthrough, whereas at 40$^\circ$C and 60$^\circ$C, crystals appear after only 3~min and 1~min, respectively—a 1--2 order of magnitude reduction in nucleation time. This acceleration reflects the combined effects of faster evaporation (higher vapor pressure at elevated temperature) and enhanced diffusive transport (higher diffusion coefficients). The second column shows complete dryout states, with total evaporation times of 284~min (20$^\circ$C), 180~min (40$^\circ$C), and 97~min (60$^\circ$C). The threefold reduction from 20$^\circ$C to 60$^\circ$C aligns with the exponential temperature dependence of water vapor pressure (Clausius--Clapeyron relation), which governs the thermodynamic driving force for evaporation.

The third column presents spatiotemporal evaporation maps. At 20$^\circ$C (liquid CO$_2$), evaporation is highly heterogeneous, with persistent late-stage brine pools (red regions) indicating slow, diffusion-limited drying. At 40$^\circ$C and 60$^\circ$C (gaseous CO$_2$), the dominance of blue and green regions indicates rapid, relatively uniform evaporation with minimal late-stage residues. This shift from heterogeneous to uniform drying reflects the transition from diffusion-dominated (liquid) to convection-enhanced (gas) mass transfer, consistent with the higher Sherwood numbers observed for gas-phase conditions (discussed in the following section).

The fourth column in Figure~\ref{fig:TemperatureImpact} shows crystal growth maps. At 20$^\circ$C, crystal precipitation is sparse and slow, with red regions (early crystals) dominant, reflecting limited supersaturation development due to slow evaporation. At 40$^\circ$C and 60$^\circ$C, crystal distributions are denser and more spatially extensive, with broader crystal size distributions (insets). Higher temperatures promote both faster nucleation (more crystals) and faster growth (larger crystals), resulting in overall higher crystal fractions. The increased density and broader size distributions at elevated temperatures are consistent with the higher supersaturation rates achieved under faster evaporation.

\subsection{Transport Control on Crystallization: Péclet-Sherwood-Avrami Relationships}

\begin{figure}[h!]
	\includegraphics[width=0.95\textwidth]{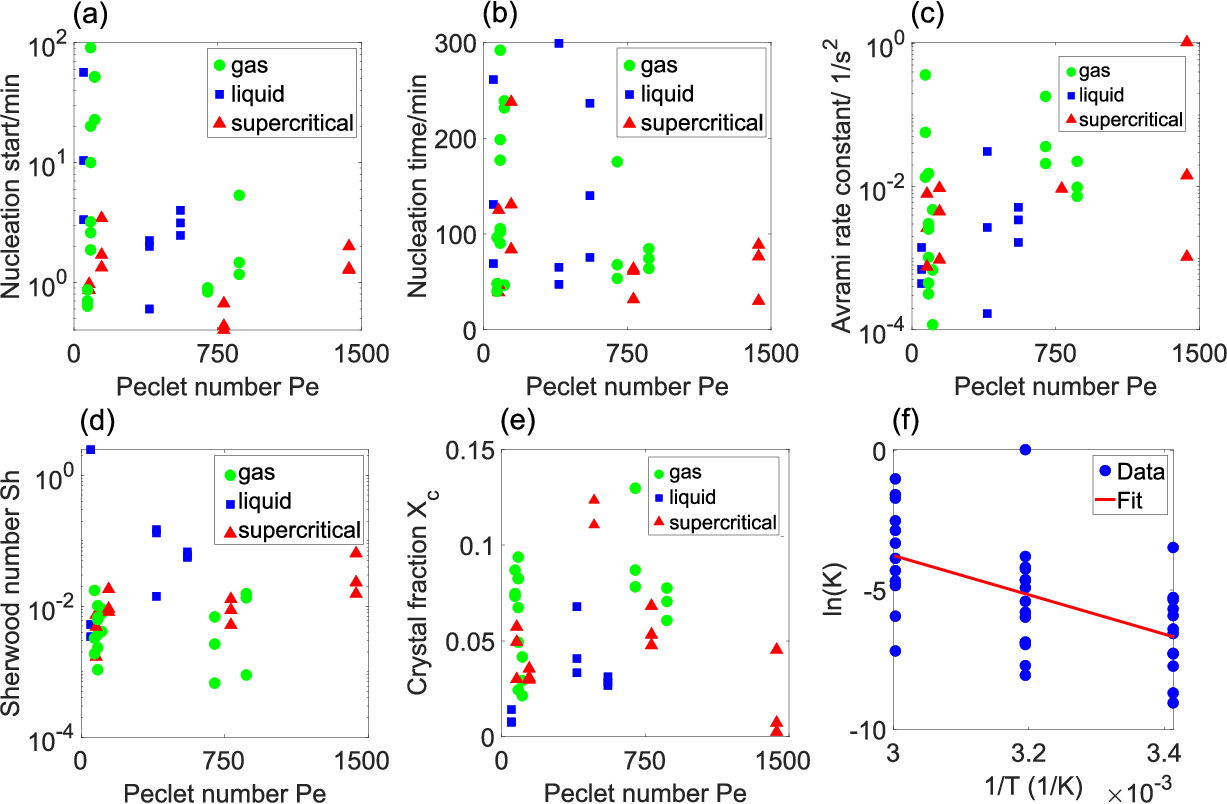}
	\caption{Transport control on crystallization kinetics across all experiments. (a) Time to first nucleation versus Péclet number. Higher Pe accelerates supersaturation, reducing nucleation time by 1--2 orders of magnitude. (b) Total precipitation-growth time versus Pe. Convection-dominated conditions (high Pe) contract nucleation duration to $<$100~min. (c) Avrami rate constant $K$ versus Pe. A nearly two-order-of-magnitude increase demonstrates transport-controlled crystallization. Supercritical CO$_2$ yields the highest $K$. (d) Sherwood number versus Pe. Enhanced mass transfer under gas/supercritical conditions confirms convection-dominated evaporation. (e) Final crystal fraction versus Pe. Generally increases with Pe, indicating more complete salt precipitation under efficient evaporation. A slight decrease at very high Pe for supercritical CO$_2$ may reflect rapid dryout, limiting solute redistribution. (f) Arrhenius plot of $\ln K$ versus $1/T$. A linear fit yields an activation energy $E_a$~=~58.6~kJ/mol, confirming thermally activated growth kinetics with intermediate control by interfacial attachment and transport processes.}	
	\label{fig:DataPointsResults}
\end{figure}

Figure~\ref{fig:DataPointsResults} synthesizes the influence of transport conditions on evaporation and crystallization kinetics across all experiments, with the Péclet number serving as the primary dimensionless parameter characterizing the convection-diffusion balance.

Figure~\ref{fig:DataPointsResults}(a) shows that the time to first nucleation (first instance of crystallite identification) decreases systematically with increasing Pe for all phases, spanning from $\sim$50~min at low Pe to $<$1~min at high Pe. This trend reflects the fact that higher convective transport accelerates brine evaporation, thereby reducing the time required to reach supersaturation. The effect is most pronounced for gaseous and scCO$_2$, where nucleation occurs within minutes even at moderate Pe, whereas liquid-phase conditions exhibit nucleation times 1--2 orders of magnitude longer. This phase dependence arises from the interplay of vapor pressure (higher for gas phases), diffusivity (higher for low-density phases), and viscosity (lower in gas/supercritical phases, enhancing convective mixing). The rapid nucleation under gas/supercritical conditions suggests that mass transfer limitations are substantially reduced, enabling rapid supersaturation buildup.

Panel (b) shows that total precipitation-growth time, defined as the interval from the first crystallite until no new crystals appear, follows a similar decreasing trend with Pe. At low Pe, precipitation proceeds slowly over extended periods (several hours), consistent with diffusion-limited solute transport and gradual concentration increases in isolated brine pools. As Pe increases, halite growth duration contracts to $<$100~min for gas and supercritical phases, indicating a transition to transport-enhanced kinetics where convective mixing rapidly distributes supersaturation throughout the residual brine network.

Panel (c) presents the Avrami rate constant $K$ versus Pe, showing a strong positive correlation over nearly two orders of magnitude. The scCO$_2$ consistently yields the highest $K$ values, followed by gaseous and liquid phases. This hierarchy reflects the cumulative effects of phase-dependent diffusivity, evaporation efficiency, and hydrodynamic mixing on overall crystallization rates. The steep increase in $K$ with Pe confirms that crystallization is predominantly transport-controlled rather than kinetically limited: when convection dominates (high Pe), evaporation and supersaturation development accelerate, driving faster crystal growth. Conversely, at low Pe, diffusion bottlenecks limit supersaturation rates, slowing crystal formation. This result has important implications for near-wellbore crystallization in storage reservoirs in saline and hypersaline aquifers, where high injection velocities (high Pe) will promote rapid salt accumulation and permeability decline unless mitigated by operational strategies.

The crossplot of Sherwood number (Sh) versus Pe (Fig.~\ref{fig:DataPointsResults}d) indicates the expected correlation between convective transport and mass transfer efficiency. Sh increases systematically with Pe, with gas and supercritical phases exhibiting significantly higher values than liquid CO$_2$ at comparable Pe. The Sh--Pe relationship can be approximated by a power-law correlation of the form $\text{Sh} \propto \text{Pe}^\alpha$, with $\alpha \approx 0.4$--0.5 typical for laminar flow in porous media. Such empirical correlations are valuable for incorporating evaporation kinetics into pore-network models and continuum-scale reactive transport simulations.

Panel (e) further shows the final crystal fraction $X_c$ as a function of Pe. For most conditions, $X_c$ increases with Pe, reaching maxima of 0.08--0.12 for gaseous and scCO$_2$ at high Pe, compared to $<$0.02 for liquid CO$_2$ at low Pe. This trend is consistent with the expectation that faster evaporation (higher Sh, higher Pe) enables more complete brine removal and thus higher salt precipitation. Interestingly, at very high Pe under supercritical conditions, $X_c$ slightly decreases, possibly due to rapid dryout that limits solute redistribution and trapping before complete crystallization can occur. This non-monotonic behavior suggests an optimal Pe range for maximizing salt precipitation, though the effect is modest and may be specific to the microfluidic geometry.

Finally, panel (f) in Figure~\ref{fig:DataPointsResults} presents an Arrhenius plot of $\ln K$ versus $1/T$, yielding a linear fit with activation energy $E_a$~=~58.6~kJ/mol and pre-exponential factor $K_0$~=~3.58~$\times$~10$^7$~s$^{-3}$. The linear relationship confirms that crystal growth kinetics are thermally activated, with temperature exerting an exponential effect on $K$. The activation energy of 58.6~kJ/mol is significantly higher than typical values for diffusion-limited processes (10--20~kJ/mol) but lower than values associated with surface reaction control (80--150~kJ/mol). This intermediate value suggests that crystallization is governed by a combination of interfacial attachment kinetics and transport-mediated supersaturation delivery. The relatively high $E_a$ also implies strong temperature sensitivity: increasing temperature from 20$^\circ$C to 60$^\circ$C accelerates crystallization by more than an order of magnitude, consistent with the reductions in nucleation and total evaporation times observed in Figure~\ref{fig:TemperatureImpact}.

\section{Conclusions}

This study provides a systematic investigation of how CO$_2$ phase states, temperature, and hydrodynamic transport jointly govern halite nucleation, precipitation, and growth during CO$_2$-driven brine evaporation and crystallization in porous media. Using high-resolution microfluidic experiments spanning liquid, gaseous, and supercritical CO$_2$ across 20--60$^\circ$C and Péclet numbers of 50--1440, we quantify displacement efficiency, evaporation rates, nucleation kinetics, and crystal growth dynamics under conditions representative of near-wellbore regions in saline and hypersaline aquifer reservoirs.

Supercritical CO$_2$ achieves superior displacement performance, with post-breakthrough brine saturations of 0.22--0.36 and fractal dimensions approaching 1.82, compared to liquid-phase saturations of 0.40--0.62 and fractal dimensions of 1.70--1.80. This enhanced connectivity directly reduces residual brine retention and minimizes heterogeneous trapping. Crystallization kinetics are transport-controlled, with the Avrami rate constant increasing by nearly two orders of magnitude from low to high Péclet numbers, demonstrating that convective mass transfer dominates over diffusion-limited processes. Temperature exerts exponential control via an Arrhenius activation energy of 58.6~kJ/mol, reducing nucleation times from 57~min at 20$^\circ$C to $<$1~min at 60$^\circ$C and accelerating total evaporation by a factor of three.

These quantitative relationships among dimensionless transport parameters (Pe, Sh) and kinetic constants ($K$, $E_a$) establish critical benchmarks for validating pore-network models, lattice-Boltzmann simulations, and reactive transport codes that predict near-wellbore permeability impairment. While the two-dimensional microfluidic geometry limits direct extrapolation to three-dimensional reservoir conditions, the mechanistic insights and scaling laws developed here provide a foundation for upscaling efforts. Operationally, the strong coupling among flow rate, reactive transport, and halite precipitation dynamics suggests that controlling injection velocities can control salt accumulation and sustain long-term injectivity in geological CO$_2$ storage.

\begin{acknowledgement}
This work was supported by the project ``Solid and Salt Precipitation Kinetics During CO$_2$ Injection into the Reservoir,'' funded by the Norway Grants (Norwegian Financial Mechanism 2014--2021) under grant number UMO-2019/34/H/ST10/00564. Additional support was provided by the EEA and Norway Grants under grant number NOR/PONORCCS/Agastor/0008/2019-00. M.N. and M.M. further acknowledge funding from the Norwegian Centennial Chair (NOCC) programme for the project ``Understanding Coupled Mineral Dissolution and Precipitation in Reactive Subsurface Environments.''
\end{acknowledgement}


\bibliography{mybib}

\end{document}